\documentclass[preprint,aps]{revtex4}
\usepackage{graphicx}
\usepackage{epstopdf}
\usepackage{dcolumn}
\usepackage{bm}

\begin{document}


\title{On the physisorption of water on graphene: a CCSD(T) study}

\author{Elena Voloshina,$^{1,}$\footnote{Corresponding author. E-mail: elena.voloshina@fu-berlin.de} Denis Usvyat,$^{2}$ Martin Sch\"utz,$^{2}$ Yuriy Dedkov,$^3$ and Beate Paulus$^{1}$}

\affiliation{$^1$Institut f\"ur Chemie und Biochemie -- Physikalische und Theoretische Chemie, Freie Universit\"at Berlin, Takustrasse 3, 14195 Berlin, Germany}
\affiliation{$^2$Institut f\"ur Physikalische und Theoretische Chemie, Universit\"at Regensburg, 93040 Regensburg, Germany}
\affiliation{$^3$Fritz-Haber-Institut der Max-Planck-Gesellschaft, 14195 Berlin, Germany}

\date{\today}

\begin{abstract}
The electronic structure of the zero-gap two-dimensional graphene has a charge neutrality point exactly at the Fermi level that limits the practical application of this material. There are several ways to modify the Fermi-level-region of graphene, e.g. adsorption of graphene on different substrates or different molecules on its surface. In all cases the so-called dispersion or van der Waals interactions can play a crucial role in the mechanism, which describes the modification of electronic structure of graphene. The adsorption of water on graphene is not very accurately reproduced in the standard density functional theory (DFT) calculations and highly-accurate quantum-chemical treatments are required. A possibility to apply wavefunction-based methods to extended systems is the use of local correlation schemes. The adsorption energies obtained in the present work by means of CCSD(T) are much higher in magnitude than the values calculated with standard DFT functional although they agree that physisorption is observed. The obtained results are compared with the values available in literature for binding of water on the graphene-like substrates. 
\end{abstract}


\maketitle

Graphene, a two-dimensional layer of carbon atoms packed in a honeycomb lattice, is a unique physical object and it is under intent attention for the last several years due to its fascinating properties.~\cite{Geim:2009} Starting from the first experimental works on the observation of ambipolar field effect and quantum Hall effect in graphene,~\cite{Novoselov:2005} this material still continues to astonish scientists, demonstrating various interesting phenomena, like high carrier mobility, integer and half-integer quantum Hall effect, Klein tunneling, etc.~\cite{Geim:2009} Furthermore, many interesting practical applications of this material have been proposed. Among them are high-frequency field-effect transistor,~\cite{Liao:2010} flexible touch-screens,~\cite{Bae:2010} single-molecule gas sensors~\cite{Schedin:2007} and many others. Almost all graphene-based devices, which are presently built or will be implemented in future, are based on the fact that in the neutral state of graphene the density of states at the Fermi level is zero and can be easily changed upon particular conditions, leading to the dramatic variation of the conductance response of the graphene conductive channel. Such conditions can be realized in different ways, like in the field-effect transistor via application of different-sign voltages to the back-gate electrode (controllable change of the conductivity of graphene from $n$- to $p$-type), adsorption of graphene on different substrates (e.\,g. graphene ribbon can connect two different metal contacts which induce different types of graphene doping,~\cite{Khomyakov:2009} thus allowing for formation of $n-p$ junctions in graphene), or via adsorption of atoms or molecules with different electron affinity on graphene (as was demonstrated in Ref.~\cite{Schedin:2007} graphene can be used for constructing precise gas sensors). In the last two cases the strength of the additional interactions on graphene, when adsorbed on a substrate or when adsorbing molecules, defines the changes in the density of states of graphene around the Fermi level and has to be carefully examined. 

Contributions to the binding of molecules on surfaces can be either chemical or physical in nature, i.\,e. chemisorption or physisorption, respectively.~\cite{Hudson:1998} Chemical binding typically implies a change in the electronic structure of both the molecule and the surface, either due to an ionic interaction, through charge transfer between substrate and adsorbate, or due to a covalent interaction, where orbitals deriving from the adsorbate and the substrate form new bonding and anti-bonding linear combinations. Physisorption, on the other hand, can arise through interaction of the permanent surface dipole with a permanent molecular dipole if it exists, through the interaction of the permanent dipole with an induced dipole or through interaction between fluctuating dipoles both in the adsorbate and the substrate. The latter contribution relates to the dispersion or van der Waals (vdW) interactions. Although, the typical binding energy of physisorption is small ($50-200$\,meV vs. more than $500$\,meV for chemisorption), this interaction plays an important role in nature and technique.~\cite{Geim:2003} 

Contrary to the local chemical interaction, the dispersion interactions originate from long-range electron correlation effects and they are not captured by the standard density-functional theory (DFT) because of the local character of commonly used  functionals. Consequently, DFT often fails to describe physisorption correctly. It is possible to improve the result when combining DFT with empirical forms for the van der Waals interaction~\cite{dft2} or when modifying existing functionals.~\cite{func1} A relatively high accuracy is accessible with a new exchange-correlation functional named van der Waals-density functional (vdW-DF), recently developed by Dion \textit{et al.}~\cite{func2} However, taking into account the lack of systematic improvability within the DFT framework, a better way would be to employ methods beyond the DFT approach.

For a correct and consistent treatment of physisorption interactions it is necessary to use high-level wave-function-based post-Hartree-Fock methods like the M\o ller-Plesset perturbation theory~\cite{MP2} or the coupled-cluster method (CC).~\cite{Cizek:1969} One problem here is that a very accurate treatment, e.\,g. with the CC method, scales very unfavorably with the number of electrons in the system. From a physical point of view, however, this difficulty should be avoidable because the correlation hole around an electron is a fairly local object. For solids it is reasonable, therefore, to transform the extended Bloch orbitals of the periodic system to localized Wannier orbitals. The reformulation of the many-body wavefunction in localized orbitals defines the group of the so-called local correlation methods. One method of this type is the method of increments, originally proposed by Stoll~\cite{stoll} and further developed by Paulus and coworkers (for Reviews, see Refs.~\cite{Paulus:2006,met-rev2}).  In this approach, a periodic HF calculation is followed by a many-body expansion of the correlation energy, where the individual units of the expansion are either atoms or other domains of localized molecular orbitals. Another local approach to the correlations problem, firstly formulated for molecules,~\cite{pulay} has been recently extended to periodic calculations. The latter has been implemented in a post-HF local-correlation computation code - CRYSCOR.~\cite{CRYSCOR} Present time this code allows for inclusion of the correlation effects at the local M\o ller-Plesset perturbation theory at second order (LMP2) level and only for non-conducting systems. Therefore, at the present level of progress, if coupled-cluster energies are desirable the quantum-chemical treatment for periodic systems is usually done using finite embedded clusters via application of the method of increments, as is also done in the present work. At the same time, at the MP2 level we essentially employ both techniques. 

Many theoretical studies have been focused on the investigation of physisorption on graphene-like substrates (for a Review, see Ref.~\cite{boukh}), several of them are devoted to the $\mathrm{H_2O}$/graphene system.~\cite{h1,h2,h4,wehling2,h2o-nh3,ribeiro,leenaerts} While \textit{ab initio} data available in literature are limited to interaction energies calculated for polycyclic aromatic hydrocarbons and range from $-104$ to $-249$\,meV, DFT studies, employing periodic approach, often give controversial results ($-1.94$\,eV~\cite{ribeiro} vs. from $-18$ to $-47$\,meV~\cite{leenaerts,wehling2}). This gives a hint, that the physisorption of water on graphene may not be reliably reproduced in DFT calculations and a more accurate quantum-chemical treatment is required. Here, we apply the CCSD(T) approach in the framework of the above method of increments to the adsorption of $\mathrm{H_2O}$ on graphene.   

The structure of this paper is as follows: The next two sections describe two local correlation methods employed in our studies (Sec.~\ref{sec:methods}) and the computational procedure (Sec.~\ref{sec:details}). Sec.~\ref{sec:results} presents the main results. Conclusions are drawn in Sec.~\ref{sec:conclu}.

\section{Methods}
\label{sec:methods}

\subsection{Method of increments for adsorption energies}

The method of increments combines Hartree-Fock (HF) calculations for periodic systems with correlation calculations for finite embedded clusters, and the total correlation energy per unit cell of a solid is written as a cumulant expansion in terms of contributions from localized orbital groups of increasing size. A detailed description of this approach can be found in Ref.~\cite{Paulus:2006} In this section we outline briefly how this method can be applied for the calculation of adsorption energies.
 
To quantify the molecule-surface interaction we define the adsorption energy as 
\begin{equation}\label{e-int}
E_\mathrm{ads}=E_\mathrm{Gr+H_2O}-E_\mathrm{Gr}-E_\mathrm{H_2O}=E^\mathrm{HF}_\mathrm{ads}+E^\mathrm{corr}_\mathrm{ads},
\end{equation}
where $E_\mathrm{Gr+H_2O}$ is the total energy of the $\mathrm{H_2O/graphene}$ system, and $E_\mathrm{Gr}$ and  $E_\mathrm{H_2O}$ are the energies of the fragments at the same coordinates as in the $\mathrm{H_2O/graphene}$ system corrected for the basis set superposition error (BSSE) according to the counterpoise scheme of Boys and Bernardi~\cite{bsse}.  The Hartree-Fock energy, $E^\mathrm{HF}_\mathrm{ads}$, is calculated for the periodic system in the standard way (for details, see Sec.~\ref{sec:details}), and the $E^\mathrm{corr}_\mathrm{ads}$ is calculated within the incremental expansion.~\cite{Muller:2009,Paulus:2009} For latter quantity one has to take into account all orbital groups, that change due to the interaction of the molecule with the surface. For example, the correlation contribution to the interaction energy within the adsorbed molecule, the 1-body increment, can be defined as follows: $\eta_A=\varepsilon_A^\mathrm{Gr+H_2O}-\varepsilon_A^\mathrm{free}$. As the total adsorption energy $E_\mathrm{ads}$ in expression (\ref{e-int}), also $\eta_A$ and all $\eta_i$ (the changes due to adsorption in the surface increments) terms are corrected for the BSSE. Other contributions occur in the system due to the simultaneous correlation of orbitals in groups from the molecule ($A$) and the surface ($i,j$), e.g., 2-body increment $\eta_{Ai}=\varepsilon_{Ai}-\varepsilon_{A}-\varepsilon_{i}$ and 3-body increment $\eta_{Aij}=\varepsilon_{Aij}-\varepsilon_{A}-\varepsilon_{i}-\varepsilon_{j}-\eta_{Ai}-\eta_{Aj}-\eta_{ij}$. $E^\mathrm{corr}_\mathrm{ads}$ can now be calculated as the sum of all $\eta$-terms taken with the proper weight factors (according to their occurrence in the system under consideration):
\begin{equation}
E^\mathrm{corr}_\mathrm{ads}=\eta_A + \sum_i\eta _{i}  + \sum_{i}\eta _{Ai} + \sum_{i<j}\eta _{ij} + \sum_{i<j}\eta _{Aij}+ ...\,\,.
\end{equation}

All incremental calculations are performed with the program package MOLPRO~\cite{molpro}, using the coupled cluster treatment with single and double excitations and perturbative triples [CCSD(T)].~\cite{ccsdt} Further details are summarized in Sec.~\ref{sec:details}.

\subsection{Local MP2 method}

In order to compare the results evaluated with the method of increments to the fully periodic model at the MP2 level, periodic LMP2 calculations have been performed additionally. The molecular local correlation scheme~\cite{pulay} has been recently generalized to periodic systems and implemented in the CRYSCOR code.\cite{CRYSCOR} It is based on the local representation of the occupied and virtual spaces by orthogonal localized orbitals (Wannier functions, WFs) and non-orthogonal projected atomic orbitals (PAOs), respectively. The WFs are constructed within the localization-symmetrization procedure and {\it a posteriori} symmetrized.~\cite{wf-locali} PAOs are generated in the reciprocal space by projecting the Fourier-images of atomic orbitals out of the occupied space.~\cite{TiO2}

Within the local approximation the virtual space for each orbital pair is restricted to so called orbital pair-domains,  i.e. sets of PAOs, centered spatially close to either of the two WFs. Such a truncation of the virtual space is justified by the exponential decay (in case of insulators) of the doubles amplitudes with mutual WF-PAO separation. The list of the pairs is also truncated based on the $R^{-6}$ decay of the correlation energy with inter-orbital distance. In periodic systems the energy from the missing pairs can be extrapolated to infinity by fitting the corresponding $C_6$ coefficients for orbital pairs of each type.~\cite{CRYSCOR} The restriction of the virtual space according to the local scheme might lead to $1-2$\,\% underestimation of the total correlation energy. Although this might amount to large value at the scale of relative energies, the local error is systematic and in most cases cancels almost completely, provided adequately large domains have been chosen.

The zero band gap of graphene causes severe difficulties for the LMP2 method. The zero denominators lead to divergence of the perturbative MP2 estimate of the correlation energy in graphene.~\cite{Kresse_MP2} Besides, the decay rate of the WFs is no longer exponential. However, since in the present work we study the adsorption energy, the local scheme allows for partitioning of the latter into intra-graphene, intra-adsorbate and inter-graphene-adsorbate contributions. This makes possible to completely eliminate the problematic part, namely the divergent intra-graphene correlation, from the LMP2 treatment. The three contributions to the interaction energy possess a clear physical interpretation.~\cite{LMP2_SAPT} Inter-pair energy describes the dispersive effects, while the intra-contributions mainly show the reduction in the magnitude of the correlation energy due to the compression of the electronic densities caused by the exchange repulsion. For highly polarizable systems like graphene the influence of the latter in the correlation energy is expected to be very small, as is also confirmed by incremental calculations (see Sec.~\ref{sec:results}). Therefore the intra-graphene contribution can indeed be safely omitted. 

Finally, due to a poor localization of the $\pi$-WFs of graphene the corresponding domains have to be taken sufficiently large. Test calculations showed that $30$-atom domains for such WFs provide almost converged values for the inter-pair energies. For other specifications of the periodic LMP2 calculations see Sec.~\ref{sec:details}.

\section{Computational details}
\label{sec:details}

\subsection{Structural models}

\begin{figure}[t]
\includegraphics[width=0.6\textwidth,angle=0]{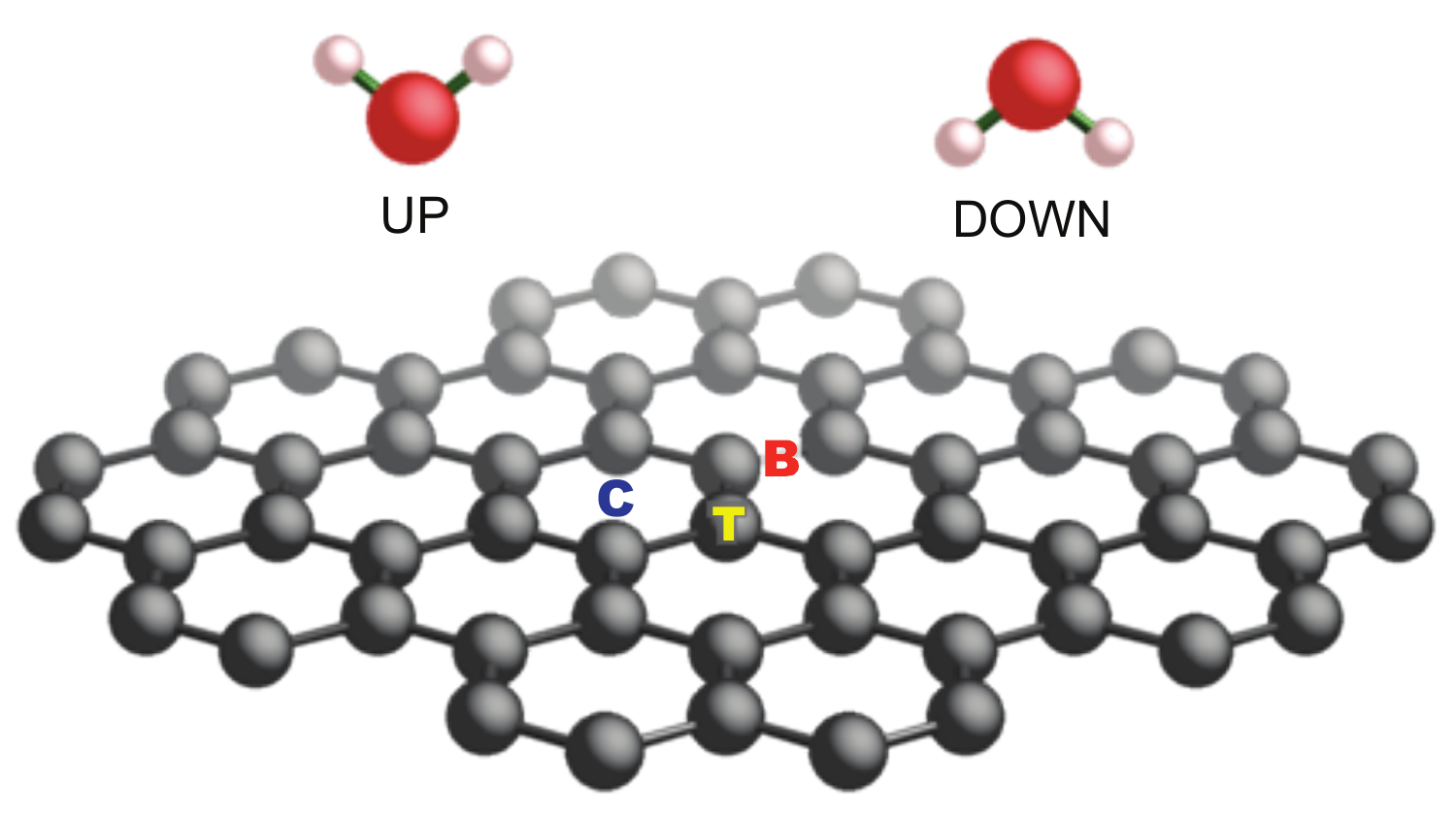}
\caption{\label{fig:structure} The $\mathrm{C_{58}H_{20}}$-cluster chosen to model the graphene sheet when performing the correlation calculations via the method of increments (hydrogens are not shown). The $\mathrm{H_2O}$-graphene arrangements are defined by the position of oxygen atom above the six-membered carbon ring (C/T/B) and the water orientation (UP/DOWN).}
\end{figure}

To model a single $\mathrm{H_2O}$ molecule on graphene, in periodic calculations $(3\times3)$ graphene supercells were used. When considering the correlation energy within the method of increments, a graphene sheet is mimicked by a finite fragment as shown in Fig.~\ref{fig:structure}. All C-C distances are set to the experimental values, i. e. $d_\mathrm{C-C}=a_\mathrm{Gr}/(2sin60^\circ)=1.421$\,\AA. The dangling bonds are saturated with hydrogen atoms and the C-H distances are set to $1.084$\,\AA. For the water geometry we used the following values: $d_\mathrm{O-H}=0.9584$\,\AA\ and $\theta_\mathrm{H-O-H}=104.45^\circ$. There was no geometry optimization performed in this work. For the water molecule, three adsorption sites are considered, namely, on top of a carbon atom (T), the center of a carbon hexagon (C), and the center of a carbon-carbon bond (B). For these positions, two different orientations of the molecule with respect to the graphene surface are examined, namely: the circumflex-like and the caron-like, denoted as UP and DOWN orientations, respectively (see Fig.~\ref{fig:structure}).

\subsection{Basis sets}

The basis sets employed are the polarized correlation-consistent valence-double-$\zeta$ basis (cc-pVDZ)~\cite{basis1} of Dunning for C and H of graphene and the aug-cc-pVTZ~\cite{basis1} basis for the water molecule. The only change needed to use these basis sets in the periodic case is the modification of $p$-type GTO from the original cc-pVDZ basis for the C atom: original $\alpha_p=0.1517\,\,\mathrm{Bohr^{-2}}$ was set to $0.17\,\,\mathrm{Bohr^{-2}}$. Apart from that, the standard molecular basis sets have been employed without any change. For test calculations we additionally employed a carbon triple-zeta-quality $[6s3p2d1f]$ basis for graphene (to be denoted below as VTZ). For this basis the $s$- and $p$-orbitals were taken from Ref.~\cite{Ahlrichs_vtz}, with a modification that the 2 smallest exponents for both $s$- and $p$- orbitals have been upscaled to $0.5$ and $0.2$ Bohr$^{-2}$. The $d$- and $f$-orbitals in this basis are those of the cc-pVTZ basis set.~\cite{basis1}

\subsection{HF and DFT calculations}

The periodic mean-field calculations for the studied system were performed with the program package CRYSTAL\,09.~\cite{crystal} In order to obtain converged results for the HF binding energy, the default parameters were modified. For the pseudo-overlap tolerances ITOL4 and ITOL5, used for the prescreening of the exchange integrals, tighter values than usual have been employed: $15$ and $80$, respectively. The other CRYSTAL tolerances (ITOL1-3), used for the screening of Coulomb integrals, have been set to $7$. The $k$ mesh shrinking factors (isotropic Monkhorst net) have also been set to $12$ corresponding to $74$ $k$-points to sample the irreducible BZ. The chosen values guarantee the stability of the HF and LMP2 solution for the considered system and basis sets. For compatibility, the DFT calculations reported here were performed with the same thresholds as used for the HF calculations. For the exchange-correlation energy functional we employed the generalized gradient approximation as parameterized by Perdew et al. (PBE).~\cite{pbe} Some additional test calculations were performed by means of the VASP code.~\cite{vasp} 

In the context of DFT, the role of the long-range van der Waals interactions on the adsorption energy of the water-graphene system was considered by employing the semiempirical approach proposed by Grimme.~\cite{dft2} This method relies on corrections added to the DFT total energy and forces, based on a damped atom-pairwise potential $C_6R^{-6}$ ($C_6$ represents the dispersion coefficient for a given atom pair and $R$ is the distance between the atoms). 

In several cases, in order to obtain an improved value for the total energy, the vdW-DF functional~\cite{func2} has been applied  to the charge 
density calculated by the VASP code. This was accomplished by utilizing the
JuNoLo-code.~\cite{junolo}

\subsection{Periodic LMP2 calculations and localization of the orbitals}

The periodic LMP2 calculations were done using the CRYSCOR code.~\cite{CRYSCOR}. The orbital domains have been specified in the following way: for the WFs of the water molecules the whole molecular units were included in the domains, in graphene for the $\sigma$-WFs the domains comprised $6$ (first and second nearest neighbors), for the $\pi$-WFs -- of $30$ atoms, respectively. Intra-graphene pairs were not included in the calculation. For the inter-graphene-water pairs the cutoff distance was set to $11$\,\AA. All the integrals have been evaluated via the density fitting approximation employing the direct space local fit and combined Poisson/Gaussian-type auxiliary basis sets of the quintuple-zeta quality.~\cite{CRYSCOR} The $20\times20$ $k$-mesh was used for generating the PAOs.~\cite{TiO2}

When employing the method of increments, we use the localization procedure according Foster and Boys~\cite{locali} as implemented in the program package MOLPRO.~\cite{molpro} The localization of the orbitals in the graphene plane is somewhat delicate. The Foster-Boys procedure yields the $\sigma$ bonds between all atoms, and localizes the $\pi$ orbitals at each second C--C bond. This causes an interesting effect of making the rings formally inequivalent to each other from the point of the number of localized $\pi$-orbitals belonging to them. Two possibilities can occur for the C-adsorption site, namely: it might correspond to (i) the center of the benzene-like ring (a combination of $3\,\sigma$ and $3\,\pi$ localized orbitals (or ''bonds'' in the standard organic chemistry description); (ii) to the center of the ring, depleted of $\pi$-localized orbitals (i.e. possessing a combination of rather $6\,\sigma$-like localized orbitals). However, our test calculations show, that final contribution to the total correlation energy is not affected by this ambiguity, due to the relatively poor localization of the $\pi$-orbitals and the weight-factor compensation. The same conclusion can be made when analyzing the results reported in Ref.~\cite{Paulus:2009}, where for the interaction energy between $\mathrm{H_2S}$ and graphene-like cluster of different sizes were used.

\section{Results and discussion}
\label{sec:results}

\begin{table*}
\small
  \caption{\ Adsorption energies ($E_\mathrm{ads}$) and equilibrium distances ($d_0$) computed for the water/graphene interface at different levels of theory. $d_0$ is defined as a distance between graphene plane and either O or H, for UP and DOWN orientations, respectively.}
  \label{tab:ccsdt}
\begin{tabular*}{\textwidth}{@{\extracolsep{\fill}}l cc cc cc cc}
    \hline
Structure &\multicolumn{2}{c}{CCSD(T)} &\multicolumn{2}{c}{MP2} & \multicolumn{2}{c}{PBE}  & \multicolumn{2}{c}{PBE-D2}  \\
                  &$d_0$ (\AA)&$E_\mathrm{ads}$ (meV)&$d_0$ (\AA)&$E_\mathrm{ads}$ (meV)&$d_0$ (\AA)&$E_\mathrm{ads}$ (meV)&$d_0$ (\AA)&$E_\mathrm{ads}$ (meV)\\
  \hline
C-UP         &$3.06$&$-108$&$3.09$&$  -98$&$3.69$&$-20$&$3.07$&$-83$ \\
C-DOWN  &$2.61$&$-123$&$2.66$&$-106$&$3.52$&$-19$&$2.60$&$-139$\\
B-UP         &$3.05$&$-102$&$3.09$&$   -99$&$3.70$&$-18$&$3.17$&$-77$ \\
B-DOWN  &$2.64$&$-118$&$2.69$&$-103$&$3.68$&$-18$&$2.67$&$-129$ \\
T-UP         &$3.06$&$-110$&$3.08$&$-101$&$3.70$&$-19$&$3.18$&$-75$ \\
T-DOWN  &$2.69$&$-135$&$2.70$&$-116$&$3.67$&$-19$&$2.65$&$-128$\\
\hline
\end{tabular*}
\end{table*}

\begin{figure}[t]
\centering
\includegraphics[width=0.8\textwidth,angle=0]{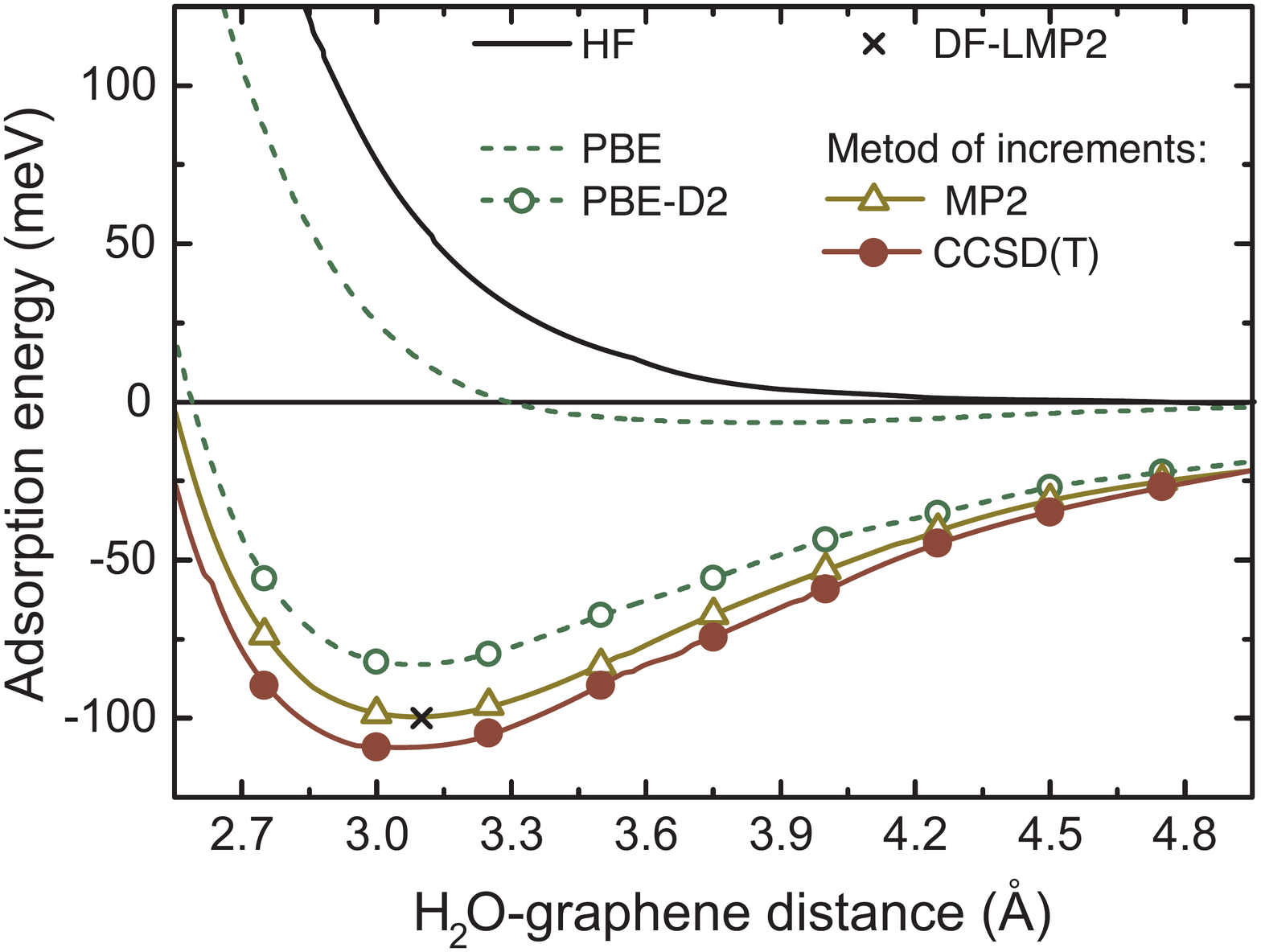}
\caption{\label{fig:diss} $\mathrm{H_2O}$-graphene interaction energy as a function of substrate-adsorbate distance as obtained with different methods for the C-UP geometry (The periodic LMP2 calculation has been done for the distance, corresponding to the minimum of the potentail curve).}
\end{figure}

\begin{figure}[t]
\centering
\includegraphics[width=0.8\textwidth,angle=0]{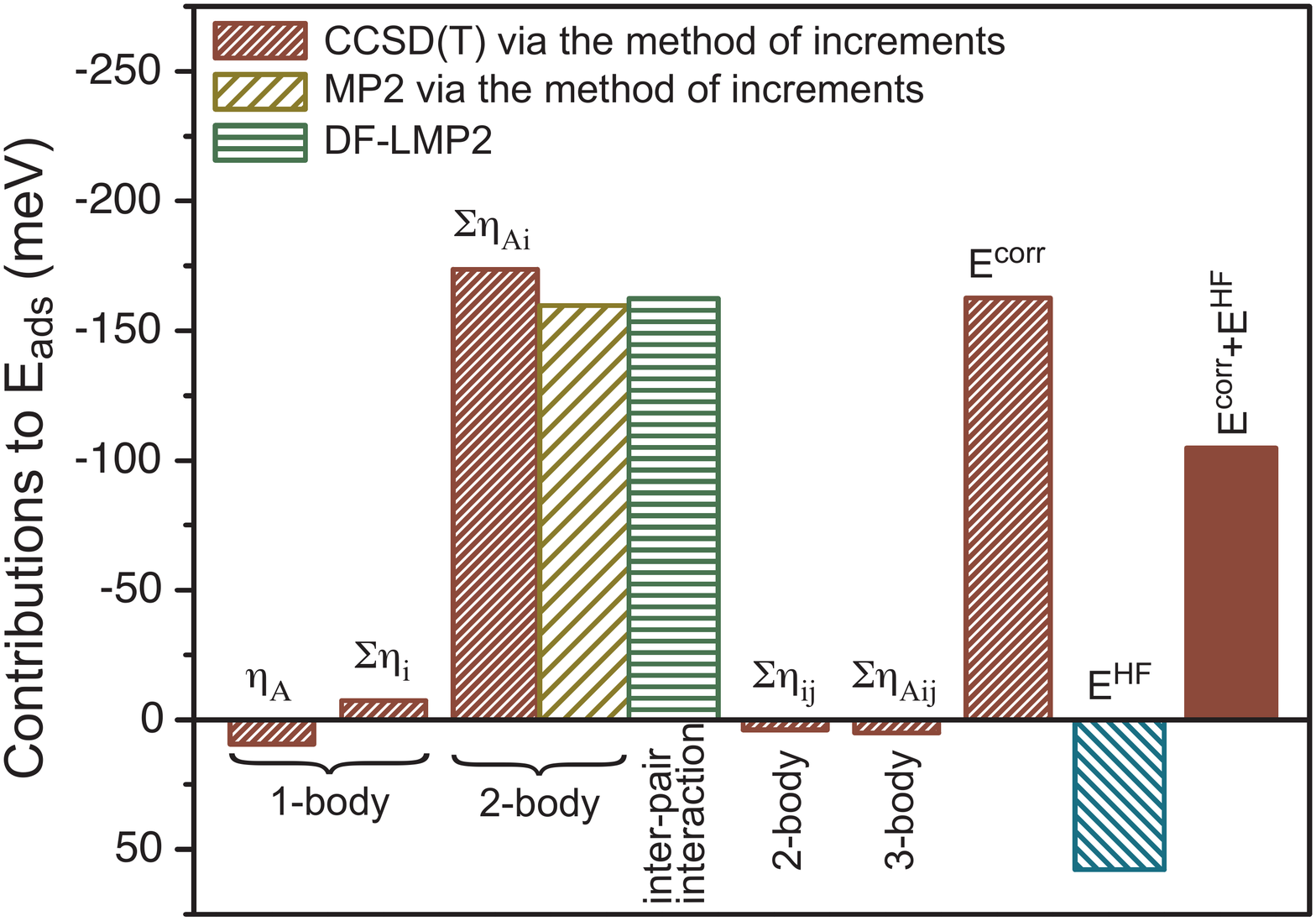}
\caption{\label{fig:fig2} The different contributions to the total interaction energy from the method of increments are plotted, for the C-UP arrangement of water molecule with respect to the graphene layer lattice when $d=3.1$\,\AA. CCSD(T) method is used when calculating the correlation-energy increments.}
\end{figure}

In Fig.~\ref{fig:diss} we plot the dissociation curve for the C-UP geometry: the HF curve is purely repulsive, and the system is stabilized by electron correlation effects. With the method of increments one gets access to the individual contributions from the different orbital groups. In Fig.~\ref{fig:fig2} the various contributions to the interaction energy of $\mathrm{H_2O}$/graphene ($d = 3.1$\,\AA) are presented.  For an estimate of the far-away contributions, which are neglected due to the cut-off the incremental expansion, we performed an $C_nR^{-n}$-fit. Thus, the correlation-energy increments for the distances up to $7.5$\,\AA\ were calculated explicitly, whereas the long-range contributions corresponding to distances up to $R=12$\,\AA\ are obtained by the fitted extrapolation. The latter brings about $3$\,\% to  $E^\mathrm{corr}_\mathrm{ads}$. While 2-body increments, implying the simultaneous correlation of orbitals in groups from the molecule and the surface ($\eta_{Ai}$), yield the major part of the interaction energy, since they describe the vdW interaction between molecule and surface, the 1-body correlated contributions are found to be small (see also below). The latter observation is because changes in the localized orbitals are already captured to a large extent at the HF level. The 2-body increments of surface, $\eta_{ij}$, and the investigated 3-body terms are even smaller, and can be neglected for further calculations. These data are in good agreement with the previously published result for the adsorption of $\mathrm{H_2S}$ on graphene.~\cite{Paulus:2009} As in that case, the largest contributions to $E_\mathrm{ads}$ originate from the $\pi$-orbitals closest to the adsorbed molecule, all $\sigma$ contributions account for only about one-quarter of the total adsorption energy.

To prove that the chosen cluster is an adequate model to describe the studied system, the periodic LMP2 calculations were performed within the same basis sets as for the incremental calculations. One observes good agreement between the periodic LMP2 and incremental MP2 correlation energies:  the intra-pair interaction, that corresponds to $\sum\eta_{Ai}$ in terms of the method of increments, obtained employing the periodic LMP2 amounts to $-162.3$\,meV while the incremental MP2 yields $\sum\eta_{Ai}=-159.7$\,meV (see Fig.~\ref{fig:fig2}). A further contribution to compare is the intra-water interaction which is equal to $9.2$\,meV and $10.4$\,meV for the periodic LMP2 and incremental MP2 calculations, respectively. The intra-graphene contribution to the interaction is indeed insignificant ($-3.4$\,meV). Therefore, the neglect of this term in the periodic calculations does not lead to a noticeable error. 

Our tests of basis set quality show that the use of the triple-zeta basis set for the graphene increases the attraction due to dispersion ($-162.3$\,meV vs. $-170.1$\,meV, for cc-pVDZ and VTZ basis sets, respectively) but at the same time increases repulsion of the HF contribution ($57.7$\,meV vs. $63.5$\,meV, respectively). Therefore, in total the increase in the interaction energy is insignificant. 

The resulting CCSD(T) as well as MP2 adsorption energies and equilibrium distances, as obtained for the six studied arrangements of the water molecule relative to the graphene layer, are listed in Tab.~\ref{tab:ccsdt}. Evidently, the DOWN orientation is clearly more preferred in this case as compared to the opposite one (i.e. UP) and the atop adsorption position is energetically most stable, although the variation in adsorption energy between different circumflex-like structures is not higher than $17$\,meV. The general trend is the same when considering the MP2-results, albeit there is a deviation in $E_\mathrm{ads}$ up to $16$\,\% as compared with the CCSD(T)-data.

It is interesting to compare the presented CCSD(T) results with the ones obtained by means of DFT. As expected, the adsorption energies evaluated with the standard PBE functional are severely underestimated. The corresponding equilibrium distances are very large and energy the minima are shallow (see e.g. Fig.~\ref{fig:diss}). These results are in good agreement with previously published data.~\cite{leenaerts} The dispersion correction term represents the dominant contribution to the binding energy. When employing the PBE-D2 scheme one finds clearly observable energy minima at reduced equilibrium distances for all the considered geometries (Tab~\ref{tab:ccsdt}). We note that the results obtained with GTO basis coincide with those evaluated with plane wave code. Whereas within the standard PBE approximation no energetic preference regarding the adsorption site or orientation of the adsorbate has been observed, dispersion corrected DFT and the CCSD(T) results agree regarding the preferable orientation of the water molecule. Moreover, the calculated equilibrium distances are very similar to each other. Surprisingly, the energy difference between the circumflex-like structure (DOWN-orientation) and its UP-counterpart is much lower when considering the CCSD(T) results. A further discrepancy between the data obtained by means of the method of increments and the PBE-D2 scheme is the preferable adsorption site: the position in the middle of the C-ring is shown to be the most stable one when using PBE-D2. To clarify the situation we performed single-point calculations for C-DOWN and T-DOWN geometries (PBE-D2 minima) employing the vdW-DF functional. As a result the T-DOWN geometry was found to be by $8$\,meV more stable than C-DOWN, being in line with the CCSD(T) values. 

When considering previous estimates of water-graphene (graphite) interactions evaluated by means of highly accurate quantum-chemical methods, one notes, that they have been scattered in the wide range from $-104$ to $-249$\,meV. The upper limit~\cite{h2} is based on an extrapolation of the MP2 interaction energies calculated for polycyclic aromatic hydrocarbons with increasing size. This interaction energy is too large in comparison to the experimental results as well as numerous simulations (see e.g.~\cite{h1} and references therein). It is known, that adsorption on graphene surface typically brings extra dispersion stabilization compared to polycyclic aromatic hydrocarbons of finite size because of an increased number of interacting carbon atoms. However, as has been shown in Ref.~\cite{h1}, water is an exception from this rule: When going from coronene  to graphene the interaction energy decreases, which can be explained by the fact that $\mathrm{H_2O}$ is a molecule with a sizable permanent dipole moment. A combination of periodic DFT (PBE approximation) and CC calculations performed for water complexes with benzene, naphthalene and coronene yields water-graphene interaction energy equal to $-135$\,meV and the equilibrium distance of about $2.68$\,\AA. Although these values seem to be in very good agreement with the result found in this work, the global minimum structure corresponds to the C-DOWN arrangement of the water molecule above the graphene surface. At the same time, the semi-empirical approach employed by McNamara \textit{et al}.~\cite{cnt} predicts T-DOWN geometry to be the most stable one, when considering complexes of a single water molecule and a single-walled carbon nanotube, even though binding is slightly overestimated as compared to our result. 

The experimental estimate for the adsorption energy of a single $\mathrm{H_2O}$ molecule on graphite surface obtained by means of microcalorimetry is $-197$\,meV.~\cite{expt} Surprisingly, the interaction energies determined directly from the experimental contact angles of water droplets on the graphite surface are shown to range from $-65$ to $-97$\,meV.~\cite{h4} The results obtained in this work cannot be directly compared to these experimental values, first of all, since further graphite layers will affect the binding due to the long-range dispersion forces. Our test calculation for graphite modeled by two-layered structure show an increase of the adsorption energy by approximately $25$\,\% (PBE-D2, C-DOWN structure). Furthermore, it may happen that the actual most stable arrangement of the $\mathrm{H_2O}$ molecule above the graphite does not coincide with any of the six structures considered here. 

\section{Conclusions}
\label{sec:conclu}

We performed CCSD(T) calculations by means of the method of increments for the adsorption of $\mathrm{H_2O}$ on graphene. It has been shown, that the circumflex-like orientation of water is more favourable than the caron-like one. Atop adsorption site is preferred by the water molecule and the most stable structure is characterized by an adsorption energy of $-135$\,eV. Qualitatively this result is reproducible at MP2 level of theory, although the water-graphene interaction is systematically underestimated as compared to the benchmark. Both CCSD(T) and MP2 yield significantly larger adsorption energies than the previously reported periodic-DFT data. This is a consequence of the local nature of the commonly used functionals (e.g. PBE) and the result can be substantially improved when applying post-DFT dispersion corrected schemes. The semiempirical PBE-D2 treatment predicts reasonable adsorption energies and equilibrium distances, yet giving at the same time some discrepancies regarding the adsorption position compared to the CCSD(T) result. These discrepancies are eliminated when applying the vdW-DF functional on top of the charge densities calculated using the PBE approximation. From Tab.~\ref{tab:ccsdt} it is clear that the water-graphene potential is very shallow, particularly so at the level of the CCSD(T) and MP2 methods. Relative stabilities of the individual minima thus depend sensitively on the choice of method. Appreciating also the challenges of a proper water simulation~\cite{h2o-struct} probably only a combination of accurate quantum-chemical calculations, DFT, and molecular dynamics can provide a reliable description of the water-graphene interface.


\section*{Acknowledgements}
We appreciate the support from the German Research Foundation (DFG) through the Collaborative Research Center (SFB) 765 ``Multivalency as chemical organisation and action principle: New architectures, functions and applications'' (EV and BP) and through the Priority Program (SPP) 1459 ``Graphene'' (YD). The authors would like to thank Professor K. Roscisewski (Krakow), and Dr. C. M\"uller (Berlin) for useful suggestions and discussions.  The computing facilities (ZEDAT) of the Freie Universit\"at Berlin  and the High Performance Computing Network of Northern Germany (HLRN) are acknowledged for computer time.

\end{document}